\documentclass[sigconf]{acmart}

\AtBeginDocument{\DeclareCaptionSubType{lstlisting}}

\usepackage{moreverb}
\usepackage{mdframed}
\usepackage{array}
\usepackage{colortbl}
\newcolumntype{L}[1]{>{\raggedright\arraybackslash}p{#1}}
\usepackage{graphicx}
\usepackage{graphics}
\usepackage{subcaption}
\usepackage{textcomp}
\usepackage{colortbl}
\usepackage{multirow}
\usepackage{multicol}
\usepackage{url}
\usepackage{booktabs}
\usepackage{framed}
\usepackage{color}
\usepackage{xspace}
\usepackage{hyperref}
\usepackage{algorithm}
\usepackage{algpseudocode}

\newcommand\BibTeX{{\rmfamily B\kern-.05em \textsc{i\kern-.025em b}\kern-.08em
T\kern-.1667em\lower.7ex\hbox{E}\kern-.125emX}}

\usepackage{lineno}

\usepackage{booktabs}
\usepackage{multirow}
\usepackage{graphicx}

\newcommand{\lstbg}[3][0pt]{{\fboxsep#1\colorbox{#2}{\strut #3}}}

\usepackage{enumitem}
\usepackage{parcolumns}
\usepackage{listings}
\lstdefinelanguage{diff}{
  basicstyle=\ttfamily\small,
  morecomment=[f][\lstbg{red!20}]-,
  morecomment=[f][\lstbg{green!20}]+,
  morecomment=[f][\textit]{@@},
  morecomment=[f][\textit]{---},
  morecomment=[f][\textit]{+++},
}

\newcommand{\eclipse}{\textsc{Eclipse}}

\newcommand{\intellij}{\textsc{IntelliJ}}
\newcommand{\saferefactor}{\textsc{SafeRefactor}}

\newcommand{\rextractorplus}{\textsc{ReExtractor+}}
\newcommand{\rminer}{\textsc{RefactoringMiner}}
\newcommand{\refdiff}{\textsc{RefDiff}}
\newcommand{\reffinder}{\textsc{Ref-Finder}}

\newcommand{\rcrawler}{\textsc{RefactoringCrawler}}

\newcommand{\jdolly}{\textsc{JDolly}}

\newcommand{\totalTransformations}{858}

\newcommand{\review}[1]{{#1\normalfont}}

\newcommand{\tool}{\textsc{RefModel}}

\newcommand{\phimodel}{\textsc{Phi4-14B}}
\newcommand{\correctphismall}{79.4\%{}} 
\newcommand{\correctphireal}{77.3\%{}} 
\newcommand{\claude}{\textsc{Claude 3.5 Sonnet}}
\newcommand{\claudefour}{\textsc{Claude 4 Sonnet}}
\newcommand{\correctclaudesmall}{98.5\%{}}
\newcommand{\correctclaudereal}{93.2\%{}} 

\newcommand{\ominihigh}{\textsc{o4-mini-high}}

\newcommand{\gemini}{\textsc{Gemini 2.5 Pro}}

\newcommand{\numTransformations}{858}

\newcommand{\numTransformationsReal}{44}
\newcommand{\numTransformationsAllReal}{64}

\microtypesetup{activate={true,nocompatibility}, patch=none}

\setcopyright{none}
\renewcommand\footnotetextcopyrightpermission[1]{}

\begin{document}

\title{\review{\tool{}:} Detecting Refactorings using Foundation Models}

\author{Pedro Simões}
 \orcid{0009-0000-2004-3591}
 \affiliation{%
   \normalsize \institution{Federal University of Campina Grande} \country{Brazil}
 }
 \email{pedro.henrique.lima.simoes@ccc.ufcg.edu.br}
 
\author{Rohit Gheyi}
 \orcid{0000-0002-5562-4449}
 \affiliation{%
   \normalsize \institution{Federal University of Campina Grande} \country{Brazil}
 }
 \email{rohit@dsc.ufcg.edu.br}

 \author{Rian Melo}
 \orcid{0009-0004-1060-7890}
 \affiliation{%
   \normalsize \institution{Federal University of Campina Grande} \country{Brazil}
 }
 \email{rian.melo@ccc.ufcg.edu.br}
 
 \author{Jonhnanthan Oliveira}
 \orcid{0000-0002-7782-410X}
 \affiliation{%
   \normalsize \institution{Federal University of Campina Grande} \country{Brazil}
 }
 \email{jonhnanthan@copin.ufcg.edu.br}

 \author{Márcio Ribeiro}
 \orcid{0000-0002-4293-4261}
 \affiliation{%
   \normalsize \institution{Federal University of Alagoas} \country{Brazil}
 }
 \email{marcio@ic.ufal.br}

 \author{Wesley K. G. Assunção}
 \orcid{0000-0002-7557-9091}
 \affiliation{%
   \normalsize \institution{North Carolina State University} 
   \country{United States}
 }
 \email{wguezas@ncsu.edu}

\begin{abstract}
Refactoring is a common software engineering practice that improves code quality without altering program behavior. 
Although tools like \rextractorplus{}, \rminer{}, and \refdiff{} have been developed to detect refactorings automatically, they rely on complex rule definitions and static analysis, making them difficult to extend and generalize to other programming languages.
In this paper, \review{we investigate the viability of using foundation models for refactoring detection, implemented in a tool named \tool{}.} We evaluate \phimodel{}, and \claude{} on a dataset of \numTransformations{} single-operation transformations applied to artificially generated Java programs, covering widely-used refactoring types. We also extend our evaluation by including \gemini{} and \ominihigh{}, assessing their performance on \numTransformationsReal{} real-world refactorings extracted from four open-source projects. These models are compared against \rminer{}, \refdiff{}, and \rextractorplus{}.
\review{\tool{} is} competitive with, and in some cases outperform, traditional tools. In real-world settings, \claude{} and \gemini{} jointly identified 97\% of all refactorings, surpassing the best-performing static-analysis-based tools. \review{The models showed encouraging generalization to Python and Golang. They provide natural language explanations and require only a single sentence to define each refactoring type.}
\end{abstract}

\maketitle

\section{Introduction}
\label{sec:introduction}

Refactoring is the process of modifying a program's internal structure to improve readability, maintainability, and design quality, while preserving its external behavior~\cite{Fowler-book-1999,Mens-TSE-2004,Opdyke-PHD-1992,Opdyke-SOOPPA-1990}. Recent studies show that more than 40\% of developers perform refactorings on a daily basis~\cite{golubev-fse-2021}.
\review{Accurate refactoring detection is essential for understanding software evolution, supporting tasks such as code review, change integration, and program comprehension, while also enabling automated adaptation of client code, data-driven refactoring tools, and empirical studies on the impact of refactoring on readability, code smells, and technical debt~\cite{refdiff-tse2020,palomba-icpc-2017,reextractor}.} However, developers have reported a lack of tool support for integrating refactorings into collaborative workflows, reviewing refactoring-specific edits, and defining new refactoring types~\cite{Kim-FSE-2012}. These limitations reduce tool adoption in software maintenance processes.

To address this, the research community has developed a variety of tools that automatically detect refactorings by analyzing version histories~\cite{Tsantalis:TSE:2020:RefactoringMiner2.0,refdiff-tse2020,reextractor}. Many of these tools are based on definitions inspired by Fowler's catalog~\cite{Fowler-book-1999}. Among the most prominent are \rextractorplus{}~\cite{reextractor}, \rminer{}~\cite{Tsantalis:TSE:2020:RefactoringMiner2.0}, and \refdiff{}~\cite{refdiff-tse2020}. \rextractorplus{} uses a context-aware statement matching algorithm for Java, \refdiff{} abstracts language-specific syntax to support JavaScript, C, and Java, and \rminer{} applies AST-based algorithms to detect refactoring types in Java, Python, and C++.
\review{Despite their value}, detecting refactorings remains a non-trivial task~\cite{4019578}. Existing approaches depend on manually crafted rules and complex static analysis techniques, making them difficult to extend or adapt to new refactoring types or programming languages. 

The emergence of foundation models offers a promising alternative by enabling code understanding through learned representations and natural language prompts~\cite{se-llms-2023,wang2024,DBLP:conf/pldi/Xu0NH22}. Fan et al.~\cite{DBLP:conf/fose-ws/FanGHLSYZ23} have highlighted the increasing use of models in computer science, and some open problems in the refactoring area. However, the extent to which these models can reliably detect refactorings, particularly in real-world software, remains underexplored.

\review{This study examines the viability of foundation models---\phimodel{} (Microsoft), \claude{} (Anthropic), \ominihigh{} (OpenAI), and \gemini{} (Google)---for detecting refactorings using our tool (\tool{}).} We evaluate these models using a dataset of \numTransformations{} transformations applied to artificially generated small Java programs, covering widely-used refactoring types. Additionally, we assess their performance on \numTransformationsReal{} refactorings extracted from four real-world software systems. For comparison, we include the results from three state-of-the-art static-analysis-based tools, namely \rextractorplus{}, \rminer{}, and \refdiff{}.
Our results show that \review{\tool{} is} competitive with traditional refactoring detection tools. For example, \phimodel{} detects \correctphismall{} of refactorings in small programs and \correctphireal{} in real systems, while \claude{} achieves \correctclaudesmall{} and \correctclaudereal{} on the same tasks, respectively. In both small and real-world settings, \claude{} and \gemini{} together correctly identified 100\% of the refactorings in the former and 97\% in the latter. The models demonstrated consistent performance across varying code sizes. In a cross-language evaluation involving 20 transformations in Python and Golang, \gemini{} detected all refactorings, while \phimodel{} detected 80\%. \claude{}, \ominihigh{} and \gemini{} outperformed the best traditional refactoring detection tools on this dataset.
All experimental data and artifacts are publicly available online~\cite{artefatos}.

These findings are encouraging, showing that foundation models not only detect a wide range of refactorings---including those missed by static-analysis-based tools---but also provide natural language explanations that can improve developer understanding. Unlike traditional tools, \review{\tool{}} only requires a single natural language sentence to define each refactoring type, simplifying extension and evolution.

\section{Evaluation: Small Programs}
\label{sec:eval-small-programs}

The goal of this evaluation~\cite{Basili1994} is to assess the effectiveness of \review{foundation models} in detecting refactorings from a developer's perspective. 
We begin with transformations applied to small programs \review{(14–34 LOC)} to better isolate and understand the behavior of the models under controlled conditions.

\subsection{Research Questions}

We address the following research questions (RQs) to achieve the goal of our study:

\begin{itemize}%
\item[RQ$_{1}$] To what extent \review{does} \phimodel{} detect refactorings? 
\item[RQ$_{2}$] To what extent \review{does} \claude{} detect refactorings? 
\item[RQ$_{3}$] How does the performance of foundation models compare to that of traditional refactoring detection tools such as \rminer{}?
\end{itemize}

\subsection{Study Design}
\label{subsec:planning-eval-small}

\review{We conducted our experiments in May 2025, using \tool{} with \phimodel{}~\cite{phi-4} and \claude{}~\cite{claude}, and assessed how varying model sizes influenced detection performance.} \phimodel{} was executed locally via the Ollama platform on a MacBook Pro equipped with an M3 processor and 18\,GB of RAM. \review{We set the temperature to 0.6~\cite{gheyi2025evaluatingeffectivenesssmalllanguage}, while keeping all other parameters at their default values.} \claude{} was accessed through its official API (\texttt{claude-3-5-sonnet-20241022}) with default settings. 
\review{All outputs from the models, both in this study and in Section~\ref{sec:eval-real-programs}, were independently assessed by a minimum of two authors. Disagreements were resolved by involving a third author.}
We also executed \rminer{} 3.0 using its default configuration.

We evaluated ten common refactoring types: Add Parameter, Encapsulate Field, Move Method, Pull Up Field, Pull Up Method, Push Down Field, Push Down Method, Rename Field, Rename Method, and Rename Class. These refactorings are widely used in practice and frequently cited in the literature~\cite{golubev-fse-2021,DBLP:conf/icse/Murphy-HillPB09}. Our evaluation includes both low-level (e.g., Rename Field) and high-level (e.g., Move Method) refactorings.

\review{We evaluate \totalTransformations{} transformations, each representing a single refactoring automatically applied—via scripting—using \eclipse{} JDT 4.16 to Java programs generated by \jdolly{}~\cite{Soares-TSE-2013,Mongiovi-TSE-2018}.}
The results produced by the foundation models and \rminer{} are compared against this known baseline.
Our choice of using \eclipse{} for applying refactorings is supported by findings from Oliveira et al.~\cite{oliveira-ist-2023}, who conducted a survey showing that in 4 out of 6 cases, developers preferred the mechanics of \eclipse{} over those used by traditional detection tools.
\review{
Our initial dataset consisted of 100 transformations for each refactoring type. We applied automated techniques to filter out incorrect transformations generated by \eclipse{} that introduced compilation errors or behavioral changes, using both the Java compiler and \textsc{SafeRefactor}~\cite{Soares-IEEE-2010,mongiovi-scp-2014}. Additionally, previous studies have shown that \eclipse{} may produce incorrect transformations~\cite{Oliveira-sbmf-2020}. For instance, when applying the Pull Up Method or Push Down Method refactorings, the tool occasionally failed to move methods to or from the appropriate superclass or subclass. Such invalid cases were filtered from our dataset using the approach proposed by Oliveira et al.~\cite{Oliveira-sbmf-2020}.
}

We use the following prompt to evaluate each transformation applied to the small Java programs. The \texttt{definitions} block refers to the refactoring descriptions used in our study, as presented in Table~\ref{tab:refactoring-definitions}. These definitions were inspired by Fowler's catalog~\cite{Fowler-book-1999,Fowler-site}.
\review{We included more refactoring types than those present in our studies to test the models' robustness and detect potential false positives. The models did not face context window constraints, as the \texttt{original} and \texttt{refactored} programs were small (14–34 LOC) and could be fully included in the prompt.}
\begin{mdframed}[backgroundcolor=cyan!5, linecolor=black, linewidth=0.5pt]
\footnotesize
\noindent You are an expert coding assistant specialized in software refactoring, with many years of experience analyzing code transformations.

\noindent You will be given two versions of a program: \\

\noindent **Original Version:** \\
\noindent \texttt{original} \\

\noindent **Transformed Version:** \\
\noindent \texttt{refactored} \\

\noindent Your task is to identify which refactoring type(s) have been applied in transforming the original program into the new version. Use only the following list of predefined refactorings: \\
\noindent \texttt{definition} \\

\noindent **Instructions:** \\
\noindent 1. Begin your response with a bullet-point list of the refactoring type(s) applied. \\
\noindent 2. Then, briefly justify each identified refactoring with reference to the specific code changes. \\
\noindent 3. Only include refactorings from the list above. \\
\noindent 4. Be concise but precise in your explanations. \\
\noindent Do not generate explanations unrelated to the given transformation.
\end{mdframed}

\begin{table}[!tp]
\centering
\footnotesize
\caption{Refactorings and their definitions.}
\label{tab:refactoring-definitions}
\rowcolors{2}{white}{gray!10}
\resizebox{\linewidth}{!}{
\begin{tabular}{|L{0.2\linewidth}|L{0.6\linewidth}|}
\hline
\rowcolor{black}
\textcolor{white}{\textbf{Refactoring}} & \textcolor{white}{\textbf{Definition}} \\
\hline
Add Met. Param. & Introduces a new parameter to an existing method. \\
Encapsulate Field & Makes a field private and adds getter and setter methods. \\
Extract Class & Moves a group of related fields and methods from an existing class into a newly created class. \\
Extract Interface & Creates a new interface from existing method(s) in a class. \\
Extract Superclass & Creates a new superclass to encapsulate shared attributes and behavior from two or more existing classes. \\
Inline Class & Merges a class into another when it is too small or redundant. \\
Inline Method & Replaces a method call with the method's body. \\
Move Field & Relocates a field to a more appropriate class. \\
Move Method & Relocates a method to a more appropriate class. \\
Pull Up Field & Moves a field from a child class to its parent class. \\
Pull Up Method & Moves a method from a child class to its parent class. \\
Push Down Field & Moves a field from a parent class to one or more child classes. \\
Push Down Method & Moves a method from a parent class to one or more child classes. \\
Remove Method Parameter & Eliminates an existing parameter from a method signature. \\
Rename Field & Changes the name of a class or instance variable. \\
Rename Method & Changes the name of a method while preserving its behavior. \\
Rename Package & Changes the name of a package declaration. \\
Rename Class & Changes the name of a class without altering its structure. \\
Rep. Magic Num. with Cons. & Replaces a literal number with a named constant. \\
\hline
\end{tabular}
}
\end{table}

\subsection{Results}
\label{subsec:results-eval-small}

Table~\ref{tab:results-small} presents the performance of three refactoring detection approaches, namely \phimodel{}, \claude{}, and \rminer{} using the metric pass@1~\cite{pass-k}. \review{The pass@1 metric measures whether the first answer produced by a model is correct.} The results indicate that both \claude{} and \rminer{} perform robustly, achieving recall values of 98.5\% and 95.8\%, respectively. In contrast, \phimodel{}, despite being a significantly smaller model, still achieves a reasonable recall of 79.4\%. For precision, \claude{} and \rminer{} maintain high precision (88.5\% and 89.3\%, respectively), whereas \phimodel{} lags behind with 56.4\%.
Certain refactoring types—such as Add Parameter, Encapsulate Field, and Rename Class—are consistently detected by all tools with high accuracy. However, more complex transformations like Move Method and Push Down Method exhibit greater variance across approaches, particularly for \phimodel{}. These results suggest that while \phimodel{} demonstrates potential, especially given its size and efficiency, larger models like \claude{} are more reliable across the evaluated transformations. 
These results suggest that while \review{\phimodel{}} can be effective for certain types of refactorings, \review{\claude{}} and traditional tools still offer more robust performance in scenarios involving class hierarchy and method relocation.
\review{The models typically provided clear explanations, referencing relevant code elements and accurately describing the corresponding refactoring.}

\begin{table}[!tp]
\centering
\footnotesize
\caption{Performance of each approach on small programs.}
\label{tab:results-small}
\rowcolors{2}{white}{gray!10}
\resizebox{\columnwidth}{!}{
\begin{tabular}{|L{0.27\linewidth}|r|r|r|r|}
\hline
\rowcolor{black}
\textcolor{white}{\textbf{Refactoring}} &
\textcolor{white}{\textbf{\textsc{Phi4}}} &
\textcolor{white}{\textbf{\textsc{Claude}}} &
\textcolor{white}{\textbf{\textsc{RMiner}}} &
\textcolor{white}{\textbf{Total}} \\
\hline
Add Parameter       &  97 & 100 & 100 & 100 \\
Encapsulate Field   & 100 & 100 & 100 & 100 \\
Move Method         &  43 &  93 & 100 & 100 \\
Pull Up Field       &  85 & 100 & 100 & 100 \\
Pull Up Method      &  37 &  45 &  45 &  47 \\
Push Down Field     &  52 & 100 & 100 & 100 \\
Push Down Method     &   0 &   9 &   3 &  11 \\
Rename Class         &  90 &  99 & 100 & 100 \\
Rename Field        &  65 &  99 & 100 & 100 \\
Rename Method       &  99 & 100 &  74 & 100 \\
\hline
\rowcolor{gray!30}
\textbf{Recall}      & \textbf{79.4\%} & \textbf{98.5\%} & \textbf{95.8\%} & \textbf{100\%} \\
\hline
\rowcolor{gray!10}
\textbf{Precision}      & \textbf{56.4\%} & \textbf{88.5\%} & \textbf{89.3\%} & \textbf{100\%} \\
\hline
\end{tabular}
}
\end{table}

\subsection{Discussion}
\label{subsec:discussion-eval-small}

Next we discuss the results of our work.

\subsubsection{Errors}
\label{subsec:discussion-eval-small-errors}

In some cases, \phimodel{} and \claude{} correctly described that a method was pushed down to a subclass. However, both models produced an invalid refactoring name for this operation, referring to it as Pull Down Member, which is not part of the refactorings presented in our prompt.
\phimodel{} also failed to distinguish between the more specific Push Down Method and Pull Up Method. Instead, it often generalized the transformation as Move Method. While this is not technically incorrect, we expected the model to recognize and name the more precise hierarchical refactoring involved.

Refactoring mechanics can be customized in various ways, as shown in previous studies~\cite{oliveira2019revisiting,DBLP:conf/icse/OliveiraAGBRGF23}. Additionally, our dataset includes several non-trivial transformation scenarios that reflect the complexity and variability encountered in real-world refactoring practices. In one case, a field was pulled from a subclass into its parent class \texttt{A}, but \texttt{A} already had a field with the same name declared in its parent class. This led \phimodel{} and \claude{} to misclassify the transformation as a Push Down Field rather than the correct Pull Up Field. Another challenging example involved pushing down an abstract method: the method was implemented in a subclass, but the abstract declaration in the superclass was not removed. Such cases introduced ambiguity for the models, as they had to infer the intended transformation from partial or inconsistent structural changes. 

\subsubsection{Prompt}
Based on preliminary results, we analyzed initial errors and refined the prompt. For example, we found that some of the definitions presented in Table~\ref{tab:refactoring-definitions} were imprecise or overly similar. In particular, the definitions for Move Method, Pull Up Method, and Push Down Method lacked clarity regarding class hierarchies. We revised them to explicitly describe the direction of transformation in the inheritance structure. Similarly, we distinguished Extract Class from Extract Superclass by emphasizing the number of classes involved and the structural context of the transformation.
To further improve prompt quality, we applied a technique known as \textit{metaprompting}~\cite{meta-prompting}, in which a foundation model is used to enhance the design of the original prompt. Specifically, we used \review{\textsc{o4-mini-high}} to revise and optimize the instructions. 

\review{
\subsubsection{\gemini{}}
We investigated whether \gemini{} could address the cases missed by \claude{} by re-running it on those transformations. Remarkably, \gemini{} correctly identified all of them.
}

\subsection{Threats to Validity}
\label{sub:small-threats}

Some validity must be considered in studies involving foundation models for software engineering tasks~\cite{sallou2024breaking}. One potential threat to internal validity concerns the accuracy and consistency of our experimental setup, particularly the design of prompts used to query the \review{foundation models}. Although we made efforts to craft prompts that were neutral and uniform across evaluations, slight variations in wording may still influence model outputs and affect comparability.
Another threat lies in our use of artificially generated small programs, created using \jdolly{}, to apply individual refactoring transformations. This strategy allows for controlled experimentation and precise isolation of each refactoring type. However, such synthetic code does not fully capture the complexity, coding conventions, or contextual dependencies present in real-world software. Consequently, the behavior of foundation models on these examples may not generalize to more realistic development environments.
To mitigate this threat, we complemented our evaluation with a dataset of refactorings applied to real-world systems, as follows.

\section{Evaluation: Real Programs}
\label{sec:eval-real-programs}

The goal of this evaluation is to assess the effectiveness of \review{foundation models} in detecting refactorings from a developer's perspective, focusing on transformations applied to real programs.

\subsection{Research Questions}

We address the following RQs to achieve the goal:

\begin{itemize}%
\item[RQ$_{1}$] To what extent \review{does \phimodel{}} detect refactorings? 
\item[RQ$_{2}$] To what extent do \claude{}, \gemini{}, \ominihigh{} detect refactorings? 
\item[RQ$_{3}$] How does the performance of foundation models compare to that of traditional refactoring detection tools such as \rextractorplus{}, \rminer{}, and \refdiff{}?
\end{itemize}

\subsection{Study Design}
\label{subsec:planning-eval-real}

\begin{table}[!tp]
\centering
\footnotesize
\caption{Java projects used for refactoring evaluation.}
\label{tab:real-projects}
\rowcolors{2}{white}{gray!10}
\resizebox{\columnwidth}{!}{
\begin{tabular}{|L{0.16\linewidth}|L{0.39\linewidth}|r|r|r|r|r|}
\hline
\rowcolor{black}
\textcolor{white}{\textbf{Project}} &
\textcolor{white}{\textbf{Domain}} &
\textcolor{white}{\textbf{KLOC}} &
\textcolor{white}{\textbf{Stars}} &
\textcolor{white}{\textbf{Contrib.}} &
\textcolor{white}{\textbf{Transf.}} \\
\hline
Lettuce & A scalable thread-safe Redis client for synchronous, asynchronous and reactive usage. & 234K & 5.6 & 135 & 30 \\
Apache Gobblin & A distributed data integration framework. & 454K & 2.6 & 115 & 1 \\
Google Maps Services & A Java client for Google Maps Services. & 38K & 1.7 & 96 & 3 \\
Spring Boot & A framework to create Spring-based applications. & 674K & 77.1 & 1,156 & 4 \\
RefMiner & A refactoring detection tool. & 127K & 0.4 & 18 & 6 \\
\hline
\end{tabular}
}
\end{table}

We conducted our experiments in May 2025\review{, using \tool{} with} \phimodel{}, \claude{}, \gemini{}~\cite{gemini}, and \ominihigh{}~\cite{o-mini}. For \phimodel{}, \claude{}, and \rminer{}, we adopted the same setup described in Section~\ref{subsec:planning-eval-small}. Additionally, \gemini{} and \ominihigh{} were manually executed via their official web interfaces using default parameters. These models are highly ranked on the Chatbot Arena leaderboard~\cite{llm-arena}. We also included \rextractorplus{} (version~2.4.4) and \refdiff{} (version~2.0) in our evaluation, both executed with their default configurations.

We evaluated \numTransformationsReal{} transformations applied to real-world open-source Java projects (see Table~\ref{tab:real-projects}) using \intellij{} (version~2024.1.4), covering 12 widely used refactoring types: Extract Class, Extract Interface, Extract Superclass, Inline Method, Move Method, Pull Up Field, Pull Up Method, Push Down Field, Push Down Method, Rename Field, Rename Method, and Rename Class. These transformations were selected based on their prevalence in real-world projects~\cite{golubev-fse-2021,DBLP:conf/icse/Murphy-HillPB09}.
\review{Each transformation consists of a single refactoring instance manually applied using \intellij{}, allowing for a more precise and controlled analysis of each approach's performance.} The results produced by the foundation models and static-analysis-based tools are compared against this baseline. 
We use the following prompt when evaluating transformations applied to \review{real} programs:

\begin{mdframed}[backgroundcolor=cyan!5, linecolor=black, linewidth=0.5pt]
\footnotesize
\noindent You are an expert coding assistant specialized in software refactoring, with many years of experience analyzing code transformations.

\noindent You will be given the diffs of a commit: \\

\noindent **Diffs:** \\
\noindent \texttt{diff}\\

\noindent Your task is to identify which refactoring type(s) have been applied in transforming the original program into the new version. Use only the following list of predefined refactorings: \\
\noindent \texttt{definition} \\

\noindent **Instructions:** \\
\noindent 1. Begin your response with a bullet-point list of the refactoring type(s) applied. \\
\noindent 2. Then, briefly justify each identified refactoring with reference to the specific code changes. \\
\noindent 3. Only include refactorings from the list above. \\
\noindent 4. Be concise but precise in your explanations. \\
\noindent Do not generate explanations unrelated to the given transformation.
\end{mdframed}

To deal with context-window~\cite{context-window} limitations, we analyze the GitHub \texttt{diff}, which includes the added, removed and some unchanged lines of code, in the prompt.
\texttt{definitions} indicates the refactoring definitions used in our work (Table~\ref{tab:refactoring-definitions}). 

\subsection{Results}
\label{subsec:results-eval-real}

Table~\ref{tab:real-java-results} presents the detection accuracy of foundation models (\phimodel{}, \claude{}, \gemini{}, \ominihigh{}) and static-analysis-based refactoring detection tools (\rextractorplus{}, \rminer{}, \refdiff{}) on a dataset of \numTransformationsReal{} using the metric pass@1~\cite{pass-k}. Each cell shows the percentage of refactorings correctly identified. The last column indicates the number of evaluated instances. Overall, \gemini{} and \ominihigh{} reached the highest recall scores, both with 93.8\%, followed closely by \claude{} (92.2\%) and \rminer{} (88.6\%). \phimodel{} obtained a recall of 79.7\%, while \refdiff{} and \rextractorplus{} reached 77.3\% and 84.1\%, respectively.
For precision, \gemini{} led with 82.2\%, followed closely by \ominihigh{} (81.1\%) and \claude{} (77.6\%). \phimodel{} reached 61.4\%, while traditional tools \rminer{} and \rextractorplus{} had lower precision values of 41.9\% and 40.2\%, respectively. \refdiff{} stood at 70.8\%. We also ran \claudefour{}, and it achieved the same results as \claude{}.
Notably, \claude{}, \gemini{}, and \ominihigh{} consistently reach perfect or near-perfect scores across most refactorings, while \phimodel{} showed strong results on simpler transformations (e.g., Rename Method) but struggled with structural ones such as Move Method and Push Down Field.

\begin{table}[!tp]
\centering
\footnotesize
\caption{\review{Performance on real-world programs.}}
\label{tab:real-java-results}
\rowcolors{2}{white}{gray!10}
\resizebox{\columnwidth}{!}{
\begin{tabular}{|L{0.22\linewidth}|r|r|r|r!{\vrule width 1.2pt}r|r|r|r|}
\hline
\rowcolor{black}
\textcolor{white}{\textbf{Refactoring}} &
\textcolor{white}{\textbf{\textsc{Phi4}}} &
\textcolor{white}{\textbf{\textsc{Claude}}} &
\textcolor{white}{\textbf{\textsc{Gemini}}} &
\textcolor{white}{\textbf{\review{\textsc{o4-mini}}}} &
\textcolor{white}{\textbf{\textsc{RMiner}}} &
\textcolor{white}{\textbf{\textsc{RefDiff}}} &
\textcolor{white}{\textbf{\textsc{ReExt.}}} &
\textcolor{white}{\textbf{Total}} \\
\hline
Extract Class       & 100\% & 0\%   & 0\%   & 0\%    & 100\% & 100\% & 100\%  & 1 \\
Extract Interface   & 100\% & 100\% & 75\%  & 100\%  & 100\% & 100\% & 75\%   & 8 \\
Extract Superclass  & 100\% & 100\% & 100\% & 100\%  & 100\% & 100\% & 100\%  & 2 \\
Inline Method       & 80\%  & 100\% & 100\% & 100\%  & 60\%  & 40\%  & 60\%   & 5 \\
Move Method         & 38\%  & 100\% & 100\% & 87.5\% & 75\%  & 100\% & 87.5\% & 8 \\
Pull Up Field       & 100\% & 100\% & 100\% & 100\%  & 100\% & 0\%   & 100\%  & 2 \\
Pull Up Method      & 80\%  & 80\%  & 80\%  & 100\%  & 100\% & 100\% & 100\%  & 5 \\
Push Down Field     & 0\%   & 100\% & 100\% & 100\%  & 100\% & 100\% & 100\%  & 1 \\
Push Down Meth.     & 50\%  & 75\%  & 100\% & 100\%  & 75\%  & 50\%  & 75\%   & 4 \\
Rename Class        & 100\% & 100\% & 100\% & 100\%  & 100\% & 100\% & 100\%  & 4 \\
Rename Field        & 100\% & 100\% & 100\% & 66.6\% & 100\% & 100\% & 100\%  & 1 \\
Rename Method       & 100\% & 100\% & 100\% & 100\%  & 100\% & 100\% & 100\%  & 3 \\
\hline
\rowcolor{gray!30}
\textbf{Recall}     & \textbf{79.7\%} & \textbf{92.2\%} & \textbf{93.8\%} & \textbf{93.8\%} & \textbf{88.6\%} & \textbf{77.3\%} & \textbf{84.1\%} & \textbf{100\%} \\
\hline
\rowcolor{gray!10}
\textbf{Precision}  & \textbf{61.4\%} & \textbf{77.6\%} & \textbf{82.2\%} & \textbf{81.1\%} & \textbf{41.9\%} & \textbf{70.8\%} & \textbf{40.2\%} & \textbf{100\%} \\
\hline
\end{tabular}
}
\end{table}

\subsection{Discussion}
\label{subsec:discussion-eval-real}

Next we discuss our results.

\subsubsection{Errors}
The types of errors observed were similar to those discussed in Section~\ref{subsec:discussion-eval-small-errors}. 
\claude{} showed recurring issues related to confusion between Pull Up and Push Down refactorings. Although it often identified the source and destination of the change correctly, it mislabeled the direction when there was no clear class hierarchy. Furthermore, \claude{} occasionally confused Move and Rename refactorings, especially when a method or field was renamed to an identifier that already existed in another class, leading it to infer relocation instead of renaming.

\gemini{} presented confusion between structurally similar refactorings. In one case, it misclassified an Extract Class as an Extract Superclass, failing to distinguish between horizontal and hierarchical decomposition. In another instance, it incorrectly described the creation of a new super-interface. Additionally, \gemini{} struggled to identify Pull Up Method refactorings when the class hierarchy was not visible in the GitHub diff. Without explicit inheritance information, the model could not infer the correct direction of the transformation.

\phimodel{} exhibited similar difficulties. One frequent error was confusing the direction of refactorings, such as using incorrect labels like ``push up'' or ``pull down'' when referring to Pull Up or Push Down Method refactorings. In transformations with multiple classes with similarly named methods, \phimodel{} often failed to correctly identify which class the method was moved from or to. Like \claude{}, it also misclassified Rename operations as Move refactorings when the new name matched an existing identifier in another class. In cases where class hierarchy information was absent, \phimodel{} similarly struggled to detect Pull Up refactorings.

\ominihigh{} was able to detect class hierarchies and correctly identify Push Down and Pull Up refactorings that were missed by \claude{} and \gemini{}, demonstrating its ability to infer structural relationships in the code. These findings underscore the importance of structural context in refactoring detection. Foundation models remain sensitive to missing or incomplete information—particularly in transformations involving class hierarchies (e.g., Pull Up and Push Down) or subtle naming conflicts (e.g., Rename vs. Move). In such cases, the absence of explicit inheritance or semantic overlap can easily lead to misclassification.

\subsubsection{Granularity}

Table~\ref{tab:accuracy-by-diff-size} presents the detection recall of various foundation models and traditional tools across different diff size (granularity) ranges, measured in LOC. Each row corresponds to a range of diff sizes, and each cell reports the percentage of correctly detected refactorings for that range. The last column indicates the total number of evaluated transformations.
The results show that most approaches, particularly \claude{}, \gemini{}, and \rminer{}, maintain high accuracy regardless of diff size, with perfect or near-perfect scores across all ranges. In contrast, \phimodel{} exhibits reduced performance for medium-sized diffs (120--159 LOC), while \refdiff{} struggles with both small and large diffs, reaching as low as 0\% in the 160--359 LOC range.

\begin{table}[!tp]
\centering
\footnotesize
\caption{Recall (\%) by diff size range (LOC) across approaches.}
\label{tab:accuracy-by-diff-size}
\rowcolors{2}{white}{gray!10}
\resizebox{\columnwidth}{!}{
\begin{tabular}{|L{0.1\linewidth}|r|r|r|r!{\vrule width 1.2pt}r|r|r|r|r|}
\hline
\rowcolor{black}
\textcolor{white}{\textbf{\textsc{Diff}}} &
\textcolor{white}{\textbf{\textsc{Phi4}}} &
\textcolor{white}{\textbf{\textsc{Claude}}} &
\textcolor{white}{\textbf{\textsc{Gemini}}} &
\textcolor{white}{\textbf{\review{\textsc{o4-mini}}}} &
\textcolor{white}{\textbf{\textsc{RMiner}}} &
\textcolor{white}{\textbf{\textsc{RefDiff}}} &
\textcolor{white}{\textbf{\textsc{ReExt.}}} &
\textcolor{white}{\textbf{Total}} \\
\hline
0--39    & 80\%  & 100\% & 100\% & 100\% & 90\%  & 30\%  & 70\% & 10 \\
40--79   & 81\%  & 86\%  & 81\%  & 86\%  & 86\%  & 76\%  & 86\% & 21 \\
80--119  & 83\%  & 100\% & 100\% & 100\% & 100\% & 67\%  & 100\% & 6 \\
120--159 & 25\%  & 100\% & 100\% & 100\% & 75\%  & 25\%  & 75\% & 4 \\
160--359 & 100\% & 100\% & 100\% & 100\% & 100\% & 0\%   & 100\% & 3 \\
\hline
\end{tabular}
}
\end{table}

\subsubsection{Evolution}
Evolving our tool to refine existing refactoring definitions or to support new refactoring types requires only updating or adding natural language definitions---such as those shown in Table~\ref{tab:refactoring-definitions}. This process is significantly simpler and more flexible than implementing the complex rules and static analysis algorithms required by traditional tools~\cite{Tsantalis:TSE:2020:RefactoringMiner2.0,refdiff-tse2020,Prete-ICSM-2010,Negara-ECOOP-2013,Dig-ECOOP-2006}. Furthermore, it enables interactive exploration of refactorings, allowing developers to query, verify, or refine the detected changes.

\subsubsection{Larger Programs}

Foundation models still face limitations related to context window length~\cite{context-window}. To address this, we designed a prompt that focuses specifically on the code transformation—rather than the entire program—enabling the models to operate effectively within the available context. \review{Although the largest \texttt{diff} in our dataset contained 359 lines, it remained within the context window limits of the evaluated models.} Our results show that foundation models can detect most refactorings and perform on par with, or slightly better than, the best static-analysis-based tools. Notably, foundation models are evolving rapidly, with newer versions offering substantially larger context windows, further expanding their applicability to software engineering tasks.
As future work, we plan to investigate the use of retrieval-augmented generation (RAG) to further enhance performance in large codebases, following recent work by Batole et al.~\cite{batole2025leveragingllmsidessemantic} that applied this strategy to automate the Move Method refactoring.

\subsubsection{Other Languages}
As a feasibility study, we extended our evaluation to include programming languages beyond Java. Using the same experimental setup, we applied 20 refactorings to three open-source projects, namely \texttt{zap}, \texttt{examples}, and \texttt{full-stack-fastapi-template}, covering 10 transformations in Python and 10 in Golang. The selected transformations produced git diffs of up to 149 lines of code. Refactorings were applied using the PyCharm 2025.1.1.1 and IntelliJ 2024.1.4 IDEs, depending on the language and project. 
\review{\phimodel{} detected 80\% of the transformations, while \claude{}, \ominihigh{}, and \gemini{} achieved 90\%, 95\%, and 100\%, respectively, highlighting their promising potential for multi-language refactoring detection.}

\subsection{Threats to Validity}

We acknowledge similar threats to validity as those discussed in Section~\ref{sub:small-threats}. A potential threat to internal validity concerns the correctness of the refactorings applied in our dataset. Although we used \intellij{} to automate the transformations, some refactorings resulted in compilation errors, behavioral changes, or were incorrectly applied~\cite{Oliveira-sbmf-2020}. To ensure the reliability of our ground truth, we manually validated and excluded all such faulty cases. To mitigate threats to external validity, we included \numTransformationsAllReal{} transformations from four real-world systems. However, the generalizability of our findings to larger and more diverse codebases remains a limitation. Additionally, we focused our analysis on smaller, isolated refactorings. Construct validity may also be affected by ambiguities in prompt phrasing or model interpretation. For instance, despite adopting a clear and consistent taxonomy (Table~\ref{tab:refactoring-definitions}), some misclassifications occurred due to non-standard terminology or overly generic responses from the models. Lastly, although our results show that foundation models can effectively detect refactorings, particularly in real-world scenarios, the size of our dataset may limit broader statistical conclusions. We did not conduct formal hypothesis testing; therefore, the observed performance differences should be interpreted with caution.

\section{Related Work}
\label{sec:related}

Dig et al.~\cite{Dig-ECOOP-2006} proposed an algorithm for detecting refactorings applied during component evolution. Their approach was implemented as an \eclipse{} plugin called \rcrawler{}. The tool was evaluated on three software components, ranging in size from 17 KLOC to 352 KLOC, and reached over 85\% accuracy across seven types of refactorings.
Prete et al.~\cite{Prete-ICSM-2010} developed \reffinder{}, a tool that detects refactorings using a template-based reconstruction approach. \reffinder{} is capable of identifying 63 out of the 72 refactoring types described in Fowler’s catalog~\cite{Fowler-book-1999}. 
The evaluation reported a precision of 79\% and a recall of 95\%.
Soares et al.~\cite{Soares-JSS-2013} conducted a comparative study of three refactoring detection approaches: (i) manual inspection~\cite{DBLP:conf/icse/Murphy-HillPB09}, (ii) commit message analysis~\cite{ratzinger-msr-2008}, and (iii) dynamic analysis using \saferefactor{}~\cite{Soares-IEEE-2010,mongiovi-scp-2014,Mongiovi-ICSME-2014}. Their analysis considered behavioral preservation as a criterion and revealed that \reffinder{} exhibited low precision and recall in practical settings.

Silva et al.~\cite{refdiff-tse2020} introduced \refdiff{} 2.0, a language-agnostic refactoring detection tool. It incorporates a novel algorithm based on Code Structure Trees, which abstracts away language-specific syntax, enabling support for multiple programming languages such as Java, C, and JavaScript.
Tsantalis et al.~\cite{Tsantalis:TSE:2020:RefactoringMiner2.0} developed \rminer{} 2.0, an AST-based tool that detects over 100 refactoring types in Java without requiring manually defined thresholds. Later, Alikhanifard and Tsantalis~\cite{rminer3.0} improved the tool's statement mapping capabilities, releasing \rminer{} 3.0 with enhanced accuracy.
To evaluate these tools, the authors executed \rminer{} 2.0, \textsc{GumTreeDiff}, and two versions of \refdiff{} on hundreds of commits from open-source GitHub projects, monitored over a two-month period using an existing dataset~\cite{Tsantalis-ICSE-2018}. The union of all true positives detected by at least one tool was used as ground truth. Manual validation was conducted by two authors, resulting in 7,226 confirmed refactoring instances across 40 refactoring types.
\refdiff{} also leveraged this dataset~\cite{Tsantalis-ICSE-2018} to evaluate its precision and recall, supplementing it with additional manually curated instances. 
We plan to expand our evaluation to incorporate this dataset, enabling a broader and more comprehensive analysis of model performance and detection accuracy.

Liu et al.~\cite{reextractor,DBLP:conf/kbse/LiuLNZLJ23} proposed \rextractorplus{}, a refactoring detection technique designed to identify both high-level and low-level refactorings. It leverages a reference-based entity matching algorithm that uses qualified names, implementation details, and reference information to match coarse-grained code entities across consecutive program versions. It incorporates a context-aware statement matching algorithm. An evaluation on real-world systems shows that \rextractorplus{} reduces false positives by 59.6\% and improves recall by 19.2\% over existing tools.

Leandro et al.~\cite{osmar-sbes-2022} proposed a technique for systematically testing refactoring detection tools. To evaluate their approach, they automatically applied 9,885 transformations across four real-world open-source projects using 8 refactorings supported by the \eclipse{} IDE. 
The authors reported 34 issues to the developers of \rminer{} and \refdiff{}.
Oliveira et al.~\cite{oliveira-ist-2023} conducted a survey with 53 developers from popular Java projects on GitHub to investigate whether the mechanics used by refactoring detection tools---such as \rminer{}~\cite{Tsantalis:TSE:2020:RefactoringMiner2.0} and \refdiff{}~\cite{refdiff-tse2020}---align with developer expectations in practice. The results revealed that these tools often fail to detect many refactorings that developers consider relevant. In four out of six scenarios presented, the majority of participants expressed a preference for the refactoring behavior implemented by the \eclipse{} IDE over that of the refactoring detection tools.

\review{In this work, we introduce \tool{}, which uses foundation models to detect code refactorings, unlike previous approaches based on static analysis. We evaluate it on two novel datasets—one synthetic and one real-world—not used in prior work.}  In both datasets, \tool{} performs on par with, or slightly better than, the state-of-the-art tools such as \rextractorplus{}, \rminer{}, and \refdiff{}. Foundation models like \claude{} and \gemini{} demonstrate strong potential as a flexible, language-independent alternative for supporting refactoring detection in modern software development. Our approach requires only a single natural language sentence to define each refactoring type, greatly simplifying its extension and evolution.

\section{Conclusion}
\label{sec:conclusion}

In this paper, \review{we evaluate the ability of foundation models to detect code refactorings using our tool \tool{}, and compare them against traditional static analysis tools.} Using a dataset of transformations applied to both synthetic and real-world Java programs, we assessed the capabilities of \phimodel{}, \claude{}, \ominihigh{} and \gemini{}, benchmarking them against \rminer{} and \refdiff{}. Our results show that foundation models---particularly \claude{}, \ominihigh{}, and \gemini{}---reach high recall and precision, in some cases outperforming the state-of-the-art static tools. Across the \numTransformationsAllReal{} refactorings applied to real-world systems, \claude{} and \gemini{} jointly reached a detection recall of 97\%. Despite being a smaller model, \phimodel{} also performed competitively across several refactoring types. Some misclassifications were observed, often due to limitations in the available context (e.g., missing class hierarchies) or ambiguities between similar refactoring types.

Foundation models are a viable and flexible alternative to traditional refactoring detection tools, particularly in scenarios where static analysis is difficult to apply or extend. Unlike rule-based tools, our approach requires only a single natural language definition of each refactoring type, significantly reducing the effort needed to support new transformations. As foundation models continue to evolve—with larger context windows, improved precision, and advancements such as retrieval-augmented generation—they are likely to become even more effective, interpretable, and adaptable across programming languages and development environments.
\review{
While promising, foundation models still involve trade-offs related to runtime and cost. We ran \phimodel{} locally on modest hardware, with each analysis completing in just a few seconds. In contrast, using LLMs incurs API-related expenses and requires sending code to the cloud, potentially raising concerns around latency and privacy.
}

As future work, we plan to expand our evaluation \review{by including non-refactoring transformations}, a broader set of refactoring types, and support for more programming languages. \review{We also aim to investigate refactorings interleaved with other code changes~\cite{DBLP:conf/icse/Murphy-HillPB09},} as well as alternative models, agentic approaches~\cite{melo2025agenticslmshuntingtest,anthropic-agents}, and diverse prompting strategies. In addition to applying refactorings using other IDEs, we will explore the detection of composite refactorings---those involving multiple operations~\cite{bibiano2020,ana-icsme-2021}---using retrieval-augmented generation techniques~\cite{batole2025leveragingllmsidessemantic}.

\review{
\section*{Acknowledgments}
We thank the reviewers for their feedback. This work was partially supported by CNPq (403719/2024-0, 310313/2022-8, 404825/2023-0, 443393/2023-0, 312195/2021-4), FAPESQ-PB (268/2025).
}


\end{document}